\documentclass[aps,prb,nobibnotes,twocolumn,superscriptaddress,bibliography]{revtex4-2}
\usepackage{amsfonts}
\usepackage{mathrsfs}
\usepackage{amsmath}
\usepackage{color}
\usepackage{natbib}
\usepackage{graphicx}
\usepackage{bm}
\usepackage{amssymb}
\usepackage{xspace}
\usepackage{epstopdf}
\usepackage{dcolumn}
\usepackage{longtable}
\usepackage{array}
\usepackage{makecell}
\usepackage{tabulary}
\usepackage{multirow}
\newcolumntype{C}[1]{>{\centering\arraybackslash}p{#1}}
\newcolumntype{L}[1]{>{\flushleft\arraybackslash}p{#1}}

\usepackage{multirow}
\usepackage{epsfig}
\usepackage[normalem]{ulem}
\usepackage{amsmath}
\usepackage{braket}
\usepackage{bbold}

\makeatletter

\usepackage[colorlinks=true, letterpaper=true, pdfstartview=FitV, linkcolor=blue, citecolor=blue, urlcolor=blue]{hyperref}

\makeatother
\begin{document}
\title{Mirror real Chern insulator in two and three dimensions}

\author{Yang Wang}
\affiliation{Key Lab of advanced optoelectronic quantum architecture and measurement (MOE), Beijing Key Lab of Nanophotonics $\&$ Ultrafine Optoelectronic Systems, and School of Physics, Beijing Institute of Technology, Beijing 100081, China}

\author{Chaoxi Cui}
\affiliation{Key Lab of advanced optoelectronic quantum architecture and measurement (MOE), Beijing Key Lab of Nanophotonics $\&$ Ultrafine Optoelectronic Systems, and School of Physics, Beijing Institute of Technology, Beijing 100081, China}

\author{Run-Wu Zhang}
\affiliation{Key Lab of advanced optoelectronic quantum architecture and measurement (MOE), Beijing Key Lab of Nanophotonics $\&$ Ultrafine Optoelectronic Systems, and School of Physics, Beijing Institute of Technology, Beijing 100081, China}

\author{Xiaotian Wang}
\affiliation{School of Physical Science and Technology, Southwest University, Chongqing 400715, China}
\affiliation{Institute for Superconducting and Electronic Materials (ISEM), University of Wollongong, Wollongong 2500, Australia}

\author{Zhi-Ming Yu}
\email{zhiming\_yu@bit.edu.cn}
\affiliation{Key Lab of advanced optoelectronic quantum architecture and measurement (MOE), Beijing Key Lab of Nanophotonics $\&$ Ultrafine Optoelectronic Systems, and School of Physics, Beijing Institute of Technology, Beijing 100081, China}

\author{Gui-Bin Liu}
\email{gbliu@bit.edu.cn}
\affiliation{Key Lab of advanced optoelectronic quantum architecture and measurement (MOE), Beijing Key Lab of Nanophotonics $\&$ Ultrafine Optoelectronic Systems, and School of Physics, Beijing Institute of Technology, Beijing 100081, China}

\author{Yugui Yao}
\affiliation{Key Lab of advanced optoelectronic quantum architecture and measurement (MOE), Beijing Key Lab of Nanophotonics $\&$ Ultrafine Optoelectronic Systems, and School of Physics, Beijing Institute of Technology, Beijing 100081, China}

\begin{abstract}
A real Chern insulator (RCI) featuring a real Chern number and a second-order boundary mode appears in a two-dimensional (2D) system  with the space-time inversion symmetry (${\mathcal{PT}}$).
Here, we propose a kind of RCI: mirror real Chern insulator (MRCI) which emerges from the system having additional horizontal mirror symmetry ${\cal{M}}_z$.
The MRCI  generally is  characterized by two independent  real Chern numbers, respectively defined in the two  mirror subsystems of the system.
Hence, the MRCI may host the second-order boundary modes different from the conventional RCI.
We show that for spinless systems, the definition of the MRCI is straightforward, as ${\mathcal{PT}}$ keeps each mirror subsystem invariant.
For the spinful systems with both ${\mathcal{PT}}$ and ${\cal{M}}_z$,  the  real Chern number for the total system remain well defined, as ${\cal{M}}_z{\mathcal{PT}}=C_{2z}{\cal{T}}$, and $(C_{2z}{\cal{T}})^2=1$.
However, since $C_{2z}{\cal{T}}$ exchanges the two mirror subsystems, the definition of the MRCI in spinful systems  requires the help of  projective symmetry algebra.
We also discuss the MRCIs in 3D systems, where the  MRCI is defined on certain mirror-invariant 2D planes.
Compared with its 2D counterpart, the 3D MRCI can exhibit more abundant physics when the systems have additional nonsymmorphic operators.
Several concrete MRCI models including  2D and 3D, spinless and spinful models  are constructed to further  demonstrate our ideas.
\end{abstract}

\maketitle

\section{Introduction}\label{intro}
Topological states of matters have become a focus in current physics research \cite{hasanColloquiumTopologicalInsulators2010a,qiTopologicalInsulatorsSuperconductors2011,chiuClassificationTopologicalQuantum2016,kruthoffTopologicalClassificationCrystalline2017}.
These materials are characterized by non-trivial topological index and novel boundary modes.
For example, a topological Chern insulator is  characterized by a $\mathbb{Z}$ valued  Chern number ${\cal{C}}$ and features chiral edge states on the boundary of the system, which  give rise to the quantized Hall conductivity
 \cite{bernevigTopologicalInsulatorsTopological2013,shenTopologicalInsulatorsDirac2017}.
The Chern insulator does not require symmetry protection but only appears in the systems that break time-reversal symmetry ${\cal{T}}$ \cite{chen_experimental_2009,stuhl_visualizing_2015}.
Interestingly, if the  ${\cal{T}}$-invariant systems can be divided into multiple subsystems, such as spin and mirror subsystems, and each  subsystem does not have ${\cal{T}}$, then the systems also exhibit a  $\mathbb{Z}$ valued topological index, such as spin Chern number \cite{kaneQuantumSpinHall2005a,kaneTopologicalOrderQuantum2005a,shengNondissipativeSpinHall2005,shengQuantumSpinHallEffect2006} and mirror Chern number \cite{teoSurfaceStatesTopological2008,takahashiGaplessInterfaceStates2011}.
The systems with nonzero  mirror Chern number are known  as mirror Chern insulator \cite{tanakaExperimentalRealizationTopological2012a,hsiehTopologicalCrystallineInsulators2012b}.
In mirror Chern insulator, the Chern number for the two mirror subsystems must be opposite to guarantee a vanishing total Chern number for the  system \cite{shaoSpinlessMirrorChern2023}.

Recently, the study  of the topological states has been extended to the real topological phases \cite{ahnBandTopologyLinking2018,zhaoSymmetricRealDirac2017,qianSecondorderTopologicalInsulator2021,qianSymmetricHigherorderTopological2022}, where the eigenstates of the electronic band are enforced to be real by   symmetry.
For 3D systems, the crucial symmetry for the real topology is  ${\mathcal{PT}}$ symmetry  with the requirement of  $({\mathcal{PT}})^2=1$ \cite{zhaoSymmetricRealDirac2017,ahnBandTopologyLinking2018}.
In 2D, there are two relevant operations, which are  ${\mathcal{PT}}$   with  $({\mathcal{PT}})^2=1$  or  ${{\cal{C}}_{2z}\mathcal{T}}$   with  $({{\cal{C}}_{2z}\mathcal{T}})^2=1$ \cite{ahnSymmetryRepresentationApproach2019a,ahnFailureNielsenNinomiyaTheorem2019b,zhuPhononicRealChern2022}.
The real Chern insulator (RCI) is characterized by a $\mathbb{Z}_2$ valued  number $\nu_R$, known as the real Chern number or  the second Stiefel-Whitney number \cite{zhaoSymmetricRealDirac2017,ahnBandTopologyLinking2018,nakahara_geometry_2003}.
It features a  second-order boundary mode--topological corners state in certain  corners.
Besides RCIs,  some 3D nodal-point and nodal-line semimetals  also have real topology and  their   second-order real topology is manifested in the topological hinge states \cite{zhao_unified_2016,bzdusek_robust_2017,ahnBandTopologyLinking2018,ahn_failure_2019,sheng_two-dimensional_2019,wang_boundary_2020,chen_second-order_2022}.

In this work, we  propose the existence of a previously unrecognized type of RCI, which emerges when the real states  have additional horizontal mirror symmetry ${\cal{M}}_z$.
In the ${\cal{M}}_z$-invarient  plane, the momentum-space Hamiltonian ${\cal{H}}({\bm k})$ can take a block diagonal form
\begin{eqnarray}\label{eq:1}
{\cal H}(\bm{k}) & = & \left[\begin{array}{cc}
h_{+}(\bm{k}) & \bm{0}\\
\bm{0} & h_{-}(\bm{k})
\end{array}\right],
\end{eqnarray}
where $h_{\pm}({\bm{k}})$ denotes the Hamiltonian of the mirror-even (mirror-odd) subsystem.
Then, if the $h_{\pm}({\bm{k}})$ also has  ${\mathcal{PT}}$, we can define a  real Chern number  $\nu_R^{\pm}$ for each mirror  subsystem.
Parallel to the discussion in the mirror Chern insulator \cite{teoSurfaceStatesTopological2008}, we term the systems with nontrivial either $\nu_R^{+}$ or $\nu_R^{-}$ as  mirror real Chern insulator (MRCI), to distinguish them from the conventional RCIs.
Interestingly, we find that  the two numbers $\nu_R^{+}$  and $\nu_R^{-}$ in MRCI generally are independent, as there does not exist a symmetry that connects the two subsystems.
Hence, the MRCI features a  $\mathbb{Z}_2\oplus\mathbb{Z}_2$  classification, labeled by two  integers:  $\nu_R^{+}$  and $\nu_R^{-}$.
This topological classification is completely different from that of mirror Chern insulator and the conventional RCIs.

We show that the MRCI can naturally appears in spinless systems with ${\mathcal{PT}}$ and ${\cal{M}}_z$ symmetries,  as ${\mathcal{PT}}$ keeps each mirror subsystem invariant.
However, for spinful systems, projective operators which caused by the   lattice $\mathbb{Z}_2$ gauge field \cite{zhaoProjectiveTranslationalSymmetry2020,zhaoSwitchingSpinlessSpinful2021} are required for the achievement of the MRCI.
For both spinless and spinful systems, we construct 2D and 3D lattice models to explicitly demonstrate the existence of the MRCIs.
We find that the MRCI in 3D systems can be completely different from that in 2D systems due to  additional nonsymmorphic spatial operations \cite{bradleyMathematicalTheorySymmetry2010,fang_new_2015,wang_hourglass_2016,yuEncyclopediaEmergentParticles2022,zhang_nonsymmorphic_2022}.
By revealing previously unknown topological phases, this work deepens our understanding on the real topology, and will stimulate further studies on their realization in real material systems.

The organization of this paper is as follows. In Sec. \ref{spinless}, we study the MRCI in both 2D and 3D systems without spin-orbit coupling (SOC).
The MRCIs in spinful systems are discussed in Sec. \ref{spinful}, in which the projective symmetry algebra of the $\mathbb{Z}_2$ gauge field and the nonsymmorphic spatial operations are presented in detail.
We give a discussion and conclusion in Sec. \ref{discussion}.

\section{MRCI in spinless systems}\label{spinless}
\subsection{2D MRCIs}\label{2Dspinless}
\begin{figure}[t]
	\includegraphics[width=1\columnwidth]{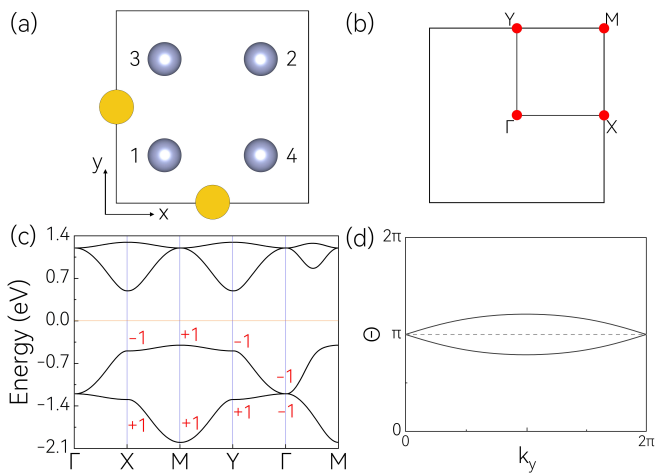}
	   \caption{(a) Atomic positions and wannier centers. The gray circle represents the atomic position 4j, and the yellow circle represents the wannier center 2e. (b) The BZ of the lattice model. (c) Calculated bulk band structure for spinless case. The parity of the TRIM is represented in red numbers. (d) Wilson loop spectrum for (1,1) direction.}
	\label{fig1}
\end{figure}

In our previous work \cite{gongHiddenRealTopology2023}, we find that the MRCI can appear in 2D ${\mathrm{Cr}}_{2}{\mathrm{Se}}_{2}\mathrm{O}$ material, in which each spin channel can be considered as an effective  spinless system.
However, a detailed  symmetry analysis  for the MRCI in spinless systems is absent there.

We begin with symmetry analysis in this section.
Consider a 2D spineless system lying in $x$-$y$ plane. This system has ${\mathcal{P}}$,  ${\mathcal{T}}$ and ${\cal{M}}_z$ symmetries with $({\mathcal{PT}})^2=1$ and $({\cal{M}}_z)^2=1$.
Besides, one knows that ${\cal{M}}_z$ commutes with both ${\mathcal{P}}$ and  ${\mathcal{T}}$, i.e. $[{\cal{M}}_z,{\mathcal{P}}]=0$ and $[{\cal{M}}_z,{\mathcal{T}}]=0$.
The total system is described by the Hamiltonian ${\cal H}({\bm k})$ ($\ref{eq:1}$), and the physics of the subsystem  with ${\cal{M}}_z=1$  (${\cal{M}}_z=-1$) is captured by $h_+$ ($h_-$).

Since the realization of the RCI require ${\mathcal{PT}}$, then for the definition of MRCI each mirror subsystem should be invarient under ${\mathcal{PT}}$.
This symmetry requirement is naturally satisfied in spineless systems, as  for eigenstates $|\psi_{\pm}\rangle$ with ${\cal{M}}_z|\psi_{\pm}\rangle=\pm 1|\psi_{\pm}\rangle$, one has

\begin{eqnarray}\label{eq:2}
{\cal{M}}_z{\cal{PT}}|\psi_{\pm}\rangle &=& {\mathcal{PT}}{\cal{M}}_z|\psi_{\pm}\rangle={\mathcal{PT}}(\pm 1|\psi_{\pm}\rangle) \nonumber \\
&=&\pm 1 ({\mathcal{PT}}|\psi_{\pm}\rangle),
\end{eqnarray}
indicating that $|\psi_{\pm}\rangle$ and ${\mathcal{PT}}|\psi_{\pm}\rangle$ share the same eigenvalue of  ${\cal{M}}_z$.
Moreover, there  does not have a symmetry that exchanges the two mirror subsystems.
Therefore, we can define two independent  $\mathbb{Z}_2$ valued real Chern numbers $\nu_R^{+}$ and $\nu_R^{-}$ for $h_+$ and $h_-$, respectively.
This means that  the MRCI features a  $\mathbb{Z}_2\oplus\mathbb{Z}_2$   classification, characterized by a topological index $\nu_M=(\nu_R^{+},\nu_R^{-})$, as discussed by Gong et. al \cite{gongHiddenRealTopology2023}.
Notice that when both mirror subsystems have nontrivial real Chern number, i.e. $\nu_M=(1,1)$,  the system would be diagnosed as trivial according to the classification of the conventional real Chern insulators \cite{zhaoSymmetricRealDirac2017,ahnBandTopologyLinking2018}.
Hence, for the system with  ${\cal{M}}_z$, one should use $\nu_M=(\nu_R^{+},\nu_R^{-})$ rather than the total real Chern number $\nu_R=\nu_R^{+}+\nu_R^{-}$ to identify its topology.

We then construct a tight-binding model with the symmetry condition specified above to explicitly show the existence of the MRCI.
We consider a 2D square lattice possessing a space group symmetry of No. 83 (P4/m), as shown in Fig. \ref{fig1}(a). Here each unit cell contains four  active site located at $4j$ Wyckoff position: $\{(\frac{1}{4},\frac{1}{4}),(\frac{3}{4},\frac{3}{4}),(\frac{1}{4},\frac{3}{4}),(\frac{3}{4},\frac{1}{4})\}$.
At each site, we put one basis orbital: $\Phi=\{d_{z^{2}}\}$.
We require that the model respects the ${\mathcal{P}}$,  ${\mathcal{T}}$ and ${\cal{M}}_z$ symmetries.
In the basis of $\Phi$, these symmetry operators take the form of
\begin{eqnarray} \label{eq:3}
{\mathcal{P}}=\Gamma_{0,1},\ && {\mathcal{T}}=\Gamma_{0,0}{\cal{K}},\   {\cal{M}}_z=\Gamma_{0,0},
\end{eqnarray}
where ${\cal{K}}$ denotes the complex conjugation operator,  $\Gamma_{i,j}\equiv \sigma_{i}\otimes\sigma_{j}$, $\sigma_{i}~(i=1,2,3)$
are Pauli matrices, and  $\sigma_{0}$ represents the $2\times 2$ identity matrix. 
Besides, for conciseness, we also impose a fourfold rotation $C_{4z}$, which is represented by
\begin{eqnarray}\label{eq:4}
C_{4z} & = & \left[\begin{array}{cccc}
0 & 0 & 0 & 1\\
0 & 0 & 1 & 0\\
1 & 0 & 0 & 0\\
0 & 1 & 0 & 0
\end{array}\right].
\end{eqnarray}

According to the  standard approach as in Refs.\cite{wieder_spin-orbit_2016,yu_quadratic_2019},  the lattice model that satisfies the above symmetries in Eq. (\ref{eq:3}) and Eq. (\ref{eq:4}) may be written as \cite{ZHANG2022108153,zhangMagneticKPPackageQuickly2023}
\begin{eqnarray}\label{eq:5}
H_{sl}^{2D}(\textbf{k}) & = &
[T_{-}\,\cos\,\frac{k_{y}}{2}\, \sin\,\frac{k_{x}}{2}-2t_{1}\,\cos \,\frac{k_{x}}{2}\,\sin\,\frac{k_{y}}{2}]\,\Gamma_{0,2}\nonumber\\
 & &
  -[T_{-}\,\cos\,\frac{k_{x}}{2}\, \sin\,\frac{k_{y}}{2}+2t_{1}\,\cos \,\frac{k_{y}}{2}\,\sin\,\frac{k_{x}}{2}]\,\Gamma_{3,2}\nonumber\\
 & &
+T_{+}\,\cos\,\frac{k_{x}}{2}\, \cos\,\frac{k_{y}}{2}\,\Gamma_{0,1}+2t_{3}\,\sin\,\frac{k_{y}}{2}\,\Gamma_{1,2} \nonumber\\
 &  & +T_{+}\,\sin\,\frac{k_{x}}{2}\, \sin\,\frac{k_{y}}{2}\,\Gamma_{3,1}+2t_{3}\,\sin\,\frac{k_{x}}{2}\,\Gamma_{2,3}  ,
\end{eqnarray}
where $T_{\pm}=t_{1}\pm t_{2}$ with $t_{1(2,3)}$ are the hopping parameters. The band structure of this model (\ref{eq:5}) with $t_{1}=0.2,\,t_{2}=1,\,t_{3}=-0.1 $ are plotted in  Fig. \ref{fig1}(c). The units of the hopping parameters here and below are all eV.
Since  all the four  basis orbital are even under ${\cal{M}}_z$, one can check that all the bands  in model (\ref{eq:5})  have the same mirror eigenvalue of ${\cal{M}}_z=1$.
This lattice model clearly demonstrates that the two mirror subsystem in spinless systems are independent, because the ${\cal{M}}_z=1$ mirror subsystem can exist independently of the ${\cal{M}}_z=-1$ mirror subsystem.

Since model (\ref{eq:5}) has both ${\cal{P}}$ and ${\cal{T}}$, we can use the real Chern number $\nu_R^+$ to describe its real topology.
Two different methods are adopted to calculate $\nu_R^+$.
The first one is counting the parity of the occupied bands at the four time-reversal invariant
points in the Brillouin zone (BZ) \cite{ahnBandTopologyLinking2018},
\begin{equation}\label{eq:6}
\ensuremath{(-1)^{\nu_R^+}=\prod_{i=1}^{4}(-1)^{\lfloor N_{\mathrm{occ}}^{-}\left(\Gamma_{i}\right)/2\rfloor }},
\end{equation}
where $\lfloor ... \rfloor$ is the floor function and $N_{\mathrm{occ}}^{-}\left(\Gamma_{i}\right)$ represents the
number of the occupied states at time-reversal invariant points $\Gamma_{i=1-4}=\{\Gamma, X, Y, M\}$ with negative inversion eigenvalues.
One  observes that the band structure undergoes twice band inversions at M point relative to   $\Gamma$ point [see Fig. \ref{fig1}(c)], which is a hallmark of non-trivial $\nu_R^+$.
Under a straightforward calculation, we obtain $\nu_R^+=1$, consistent with the analyse of the band inversion.
Since this model does not have mirror-odd bands, one has $\nu_R^-=0$ for the  mirror-odd subsystem and then   $\nu_M=(1,0)$.

The second method is using  Wilson loop  to identify the real topology of the system \cite{ahnBandTopologyLinking2018}.
For the lattice model (\ref{eq:5}), it has nontrivial Zak phase along both $(10)$ and $(01)$ directions, which can be inferred from the band  inversion at X (Y) point relative to $\Gamma$ point [see Fig. \ref{fig1}(c)].
Therefore, to correctly identify the real topology \cite{ahnBandTopologyLinking2018}, we chose the $(11)$ direction for computing
the wannier centers and the $(01)$ direction for evolution. The obtained Wilson loop is shown in Fig. \ref{fig1}(d), in which  one (odd) crossing at $\theta=\pi$ can be clearly observed.
This means that the model is nontrivial with $\nu_R^+=1$, consistent with the result of parity.

The nontrivial topology of the model (\ref{eq:5}) also can be checked  by the decomposition of the occupied  energy bands via elementary band representation \cite{bradlyn_topological_2017}.
By checking the Bilbao crystallographic server (BCS) website \cite{aroyoCrystallographyOnlineBilbao}, we find that the Wannier center of the system here does not align with the   site, but evolves from 4j to 2e Wyckoff position [see Fig. \ref{fig1}(a)], indicating a higher-order topology of this lattice model.

The lattice model (\ref{eq:5}) is  a mirror subsystem with ${\cal{M}}_{z}=1$. Similarly, we can construct a lattice model with ${\cal{M}}_{z}=-1$ by using the mirror-odd  basis orbital.
When both mirror-even and mirror-odd basis orbital are adopted, the resulted lattice models would have the bands with    ${\cal{M}}_{z}=1$ and that with ${\cal{M}}_{z}=-1$.
Then, we should divide the systems as two parts according to their eigenvalue of ${\cal{M}}_{z}$, and study the real topology of the two parts independently.

\begin{figure}[t]
	\includegraphics[width=1\columnwidth]{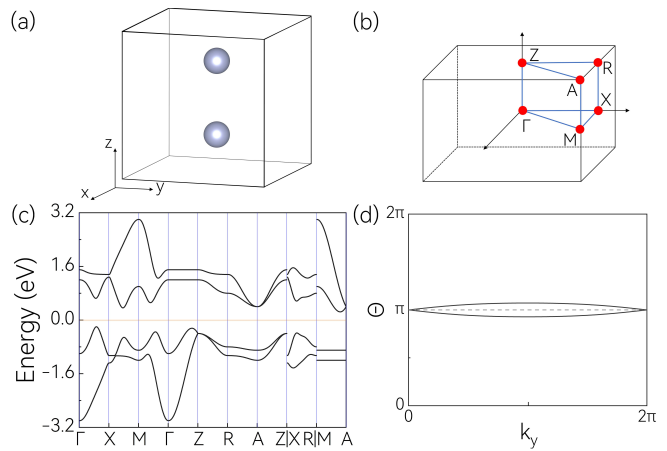}
	   \caption{(a) The view of the unit cell for the model of MRCI w/o SOC in 3D system. (b) The BZ of the lattice model (\ref{eq:12}). (c) Band structure of the lattice model (\ref{eq:12}) along high-symmetry lines. These lines are doubly degenerate. (d) Wilson loop spectrum for the plane of $k_{z}=\pi$. Only one mirror subsystem is ploted.}
	\label{fig3}
\end{figure}

\subsection{3D MRCIs}

\begin{figure}[t]
	\includegraphics[width=1\columnwidth]{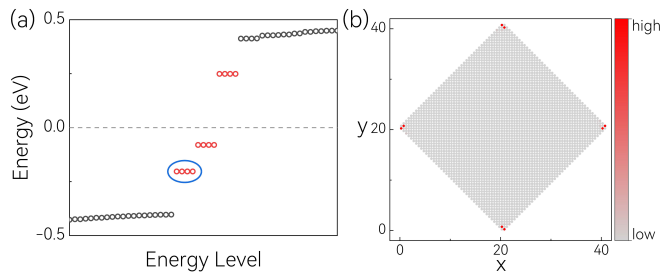}
	   \caption{(a) Energy spectrum of a $20\times20$ supercell along $(11)$ and $(1\bar{1})$ direction, and corner states are depicted by blue circle. (b) The real-space distribution of the corner states inside the bule circle in (a).}
	\label{fig2}
\end{figure}

For the  3D spinless systems  with ${\cal{M}}_z$, the MRCI  can be  defined in the ${\cal{M}}_z$-invarient planes in the BZ, i.e. the planes of $k_{z}=0$ and $k_{z}=\pi$.
Interestingly, for the space group with nonsymmorphic operators, the physics of the systems on  the BZ boundary, i.e. $k_z=\pi$ plane can be  distinct from that in a 2D systems, due to the  projective symmetry algebra.
As an example, we consider a 3D lattice possessing a space group symmetry of No. 124 (P4/mcc).
This lattice preserves ${\mathcal{P}}$,  ${\mathcal{T}}$, ${\mathcal{C}}_{4z}$,  ${\mathcal{M}}_z$ symmetries and nonsymmorphic  symmetry ${\widetilde{\mathcal{M}}}_{x}=\{{{\mathcal{M}}_{x}|00\frac{1}{2}}\}$.
Because in this lattice
\begin{eqnarray}
[{\mathcal{P}}{\mathcal{T}},{\mathcal{M}}_z]=0,
\end{eqnarray}
and ${\mathcal{M}}_z^2=1$, each  ${\mathcal{M}}_z$ subsystem in both $k_z=0$ and $k_z=\pi$ planes has ${\mathcal{P}}{\mathcal{T}}$ and well-defined real Chern number.
Note that the ${\mathcal{C}}_{4z}$ symmetry is not essential for the MRCI. It is used to simplify the Hamiltonian, thereby making the physical nature more evident.
But the nonsymmorphic symmetry ${\hat{\mathcal{M}}}_{x}$ is crucial for our discussions.
Since
\begin{equation}\label{eq:7}
{\mathcal{M}}_z {\widetilde{\mathcal{M}}}_{x}= \{E|001\}{\widetilde{\mathcal{M}}}_{x} {\mathcal{M}}_z,
\end{equation}
with $E$ the identity operator,
one has
\begin{eqnarray}
[{\mathcal{M}}_z,{\widetilde{\mathcal{M}}}_{x}]=0,
\end{eqnarray}
for $k_z=0$  plane but
\begin{eqnarray}\label{eq:9}
\{{\mathcal{M}}_z,{\widetilde{\mathcal{M}}}_{x}\}=0,
\end{eqnarray}
for $k_z=\pi$  plane.

The real topological of the $k_z=0$ is similar to that of the 2D systems with ${\mathcal{M}}_z$, classified by $\mathbb{Z}_2\oplus\mathbb{Z}_2$ index, as discussed in Sec. \ref{2Dspinless}.
Interestingly, the real topological in  $k_z=\pi$  plane would be different, due to the  nonsymmorphic  symmetry ${\widetilde{\mathcal{M}}}_{x}$.
In $k_z=\pi$  plane, for the eigenstates $|\psi_{\pm}\rangle$ with ${\mathcal{M}}_z|\psi_{\pm}\rangle=\pm 1|\psi_{\pm}\rangle$, we find that
\begin{eqnarray}\label{eq:10}
{\mathcal{M}}_z {\widetilde{\mathcal{M}}}_{x}|\psi_{\pm}\rangle &=& -{\widetilde{\mathcal{M}}}_{x} {\mathcal{M}}_z|\psi_{\pm}\rangle=\mp 1({\widetilde{\mathcal{M}}}_{x}|\psi_{\pm}\rangle) ,
\end{eqnarray}
indicating that ${\widetilde{\mathcal{M}}}_{x}$ connects the two ${\mathcal{M}}_z$  subsystems.
Consequently, the two subsystems must have the same real topology, namely, when one subsystem is trivial with $\nu_R^{\pm}=0$ (non-trivial with $\nu_R^{\pm}=1$), another is necessarily the same ($\nu_R^{+}=\nu_R^{-}$).
In such case, the $\mathbb{Z}_2\oplus\mathbb{Z}_2$ classification for the 2D MRCIs reduces to a new $\mathbb{Z}_2$ classification labeled  by $\nu_R^{+}$.
It should be noticed that this  new $\mathbb{Z}_2$ classification is completely different from the $\mathbb{Z}_2$ index defined for the generic RCIs \cite{zhaoSymmetricRealDirac2017,ahnBandTopologyLinking2018}, according to which  the $k_z=\pi$ plane is always trivial.

We then use a concrete calculation on a  lattice model to demonstrate our analysis.
As illustrated in Fig. \ref{fig3}(a), the 3D lattice with space group No. 124 is formed by stacking 2D square  lattices along the $z$ direction. Its BZ is shown in Fig. \ref{fig3}(b). Each unit cell contains two layers and two active sites located at 2b Wyckoff position: $\{(0,0,0),(0,0,\frac{1}{2})\}$. At each site, we put four basis orbitals: $\Phi=\{d_{z^{2}},d_{xy},p_{x},p_{y}\}$.
Then,  the matrix representation of the symmetries read,
\begin{eqnarray} \label{eq:11}
{\mathcal{P}}{\mathcal{T}}=\sigma_{0}\Gamma_{3,0}{\cal{K}}, {\mathcal{M}}_{z}=\sigma_{3}\Gamma_{0,0},\   {\widetilde{\mathcal{M}}}_{x}=\sigma_{1}\Gamma_{3,3}.
\end{eqnarray}
Here,  $\Gamma$ acts on the basis orbitals $\Phi$ and $\sigma$ acts on the site space.
Then, the lattice  Hamiltonian  may  be written as
\begin{eqnarray}\label{eq:12}
H_{sl}^{3D}(\textbf{k}) & = &
\chi^{+}(\textbf{k})\, \sigma_{0}(\Gamma_{0,0}-\Gamma_{3,0})+\chi^{-}(\textbf{k})\, \sigma_{0}(\Gamma_{0,3}-\Gamma_{3,3})\nonumber\\
 & &
 + 4 t_{3}\,(-\cos\,k_{x}+\cos\,k_{y})\,\sigma_{3}(\Gamma_{0,1}+\Gamma_{3,1}) \nonumber\\
 & &
 +4t_{3}\,\sin\,k_{x}U_{1}+4t_{3} \,\sin\,k_{y}U_{2}\nonumber\\
 &  &
  + 8t_{3}\,(\cos\,k_{x}+\cos\,k_{y})\, \sigma_{0}\Gamma_{3,0} \nonumber\\
 &  &
 + t_{2} \,\cos\, \frac{k_{z}}{2}\,\sigma_{2}(\Gamma_{3,2}-\Gamma_{0,2}),
\end{eqnarray}
where $\chi^{\pm}(\textbf{k})=a_{0}+t_{1}(\cos\,k_{x} \pm \cos\, k_{y})\,\cos\, k_{z}$, $U_{1}=\sigma_{0}\Gamma_{2,3}-\sigma_{0}\Gamma_{2,0}+\sigma_{3}\Gamma_{1,2}+\sigma_{3}\Gamma_{2,1}$ and $U_{2}=\sigma_{0}\Gamma_{1,2}-\sigma_{0}\Gamma_{2,1}-\sigma_{3}\Gamma_{2,0}+\sigma_{3}\Gamma_{2,3}$.
The band structure of Eq. (\ref{eq:12}) with $a_{0}=0.075,t_{1}=-0.2,t_{2}=-0.5,t_{3}=0.075$ is plotted in  Fig. \ref{fig3}(c).

On the plane of $k_{z}=0$, the  ${\mathcal{M}}_z$ eigenvalue of the eight bands are all $1$, but on the plane of $k_{z}=\pi$, the number of the bands with ${\mathcal{M}}_z=1$ and ${\mathcal{M}}_z=-1$  have to be equal, as guaranteed by Eq. (\ref{eq:9}).
When $k_{z}=\pi$, we can block diagonalize the Hamiltonian (\ref{eq:12}), according to the eigenvalue of  ${\mathcal{M}}_z$.
We calculated the  Wilson loops for  the two   ${\mathcal{M}}_z$ subsystems in  $k_{z}=\pi$ plane, and the results are shown in Fig. \ref{fig3}(d), where the Wilson loops for
 ${\mathcal{M}}_z=\pm 1$ subsystems are identical here, due to the simplicity of the lattice model.
For each ${\mathcal{M}}_z$  subsystem, the  Wilson loops exhibit an  odd number of intersections at the line of $\theta=\pi$,  showing its non-trivial topological property, i.e. $\nu_{R}^{\pm1}=1$.
For $k_{z}=0$, we also compute the Wilson loop and get $\nu_R=0$.
\begin{figure}[t]
	\includegraphics[width=1\columnwidth]{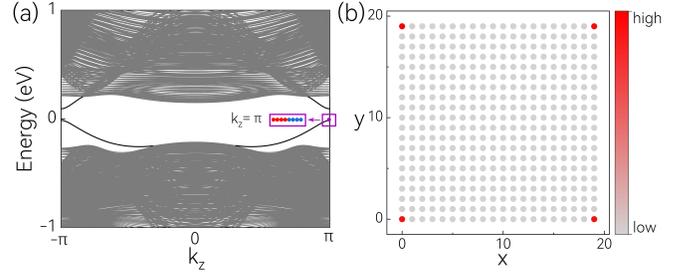}
	   \caption{(a) Hinge spectrum for model (\ref{eq:12}). The red and bule points respectively correspond to the mirror subsystems $+1$ and $-1$ on the plane of $k_{z}=\pi$. (b) The real-space distribution of the hinge states inside the purple square in (a).}
	\label{fig4}
\end{figure}

\subsection{Second-order boundary modes}
A typical feature of the real topological phases is the presence of the $(d-2)$D boundary modes on the boundary \cite{benalcazar_quantized_2017,schindler_higher-order_2018,xie_higher-order_2021}.
This  second-order boundary modes for 2D MRCIs are the corner states and for 3D  MRCIs are the hinge states \cite{langbehn_reflection-symmetric_2017,khalaf_higher-order_2018,
geier_second-order_2018,wang_higher-order_2020,ghorashi_higher-order_2020,xie_higher-order_2021}.

Here, we study the second-order boundary modes in the two lattice models (\ref{eq:5}) and (\ref{eq:12})  established above.
Since the 2D MRCI  model (\ref{eq:5}) has nontrivial real Chern number in the ${\cal{M}}_z=1$ subsystem, it must have protected corner modes at a pair of ${\mathcal{PT}}$ connected corners.
In Fig. \ref{fig2}(a), we plot the energy spectrum for the lattice  model (\ref{eq:5}) with a square geometry ($20\times20$), whose edges are in $(11)$ and $(1\bar{1})$ direction. Here, we applied an onsite energy of $-0.7$ on the edges (not including the corner sites) to remove the irrelevant edge states. We observe a set of fourfold degenerate states appearing in the gap of bulk energy bands, which is attributed to the additional presence of $C_{4z}$ symmetry.
We further visualize the spatial distribution of these states [see Fig. \ref{fig2}(b)] and find that these states primarily occupy the four corners.
This directly show  these states are corner states, confirming our topological analysis.
Moreover, one can check that these corner states have ${\cal{M}}_z=1$, as they are originated from the real topology of the ${\cal{M}}_z=1$ subsystem of the 2D MRCI  model (\ref{eq:5}).

Similarly analysis applies for the 3D MRCI model (\ref{eq:12}). Since the  $k_{z}=0$ plane is trivial, it would  not have corner state.
But for $k_{z}=\pi$ plane, it is nontrivial in both  ${\cal{M}}_z$ subsystems and should have two sets of corner states: one contributed by the ${\cal{M}}_z=-1$ subsystem  and the other by the ${\cal{M}}_z=1$ subsystem.
Based on lattice  model (\ref{eq:12}), we calculate the band structure of a rectangular crystal with surfaces in the $(100)$ and $(010)$ directions, and periodic boundary condition in the $k_{z}$ direction [see Fig. \ref{fig4}(a)]. Here, we also applied an onsite energy of $ 0.3 $ to the surfaces (not including the hinge sites).
One can  find that there are two hinge modes appearing in the  bulk band gap at $k_z=\pi$ point but disappearing at $k_z=0$ point.
Each hinge mode is eightfold degeneracy, with four connected by $\mathcal{C}_{4z}$ symmetry belonging to $\mathcal{M}_{z}=1$ and four to $\mathcal{M}_{z}=-1$, and they are connected by ${\widetilde{\mathcal{M}}}_{x}$ symmetry.
The  real-space distribution of the hinge states $k_z=\pi$ are shown in Fig. \ref{fig4}(b).
Here we can see that there are two states for each corner, and their mirror eigenvalues are $+1$ and $-1$ respectively, which is different from the conventional RCI which has only one state for each corner.

\section{MRCI in spinful systems}\label{spinful}
\subsection{2D MRCIs}\label{2Dspinful}
\begin{figure}[t]
	\includegraphics[width=1\columnwidth]{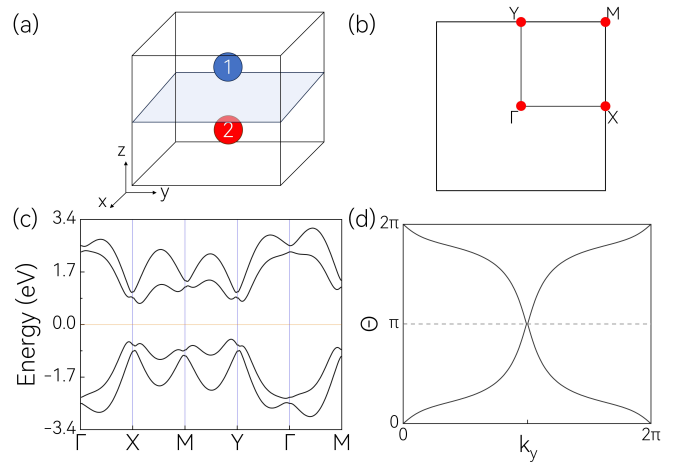}
	   \caption{(a) Two-site tight-binding model with four orbits per site. The blue and red of the atoms indicate the plus or minus as specified by $\ensuremath{\mathbb{Z}_{2}}$ gauge fields. (b) The BZ of the lattice model (\ref{eq:18}). (c) The energy bands of model (\ref{eq:18}). The bands are doubly degenerate. (d) Wilson loop spectrum for (1,0) direction. Only one mirror subsystem is ploted.}
	\label{fig5}
\end{figure}
Next, we explore the existence of the  MRCI in spinful systems.
When considering SOC, the  symmetry algebra becomes different.
For spinful  systems, one generally has  $({\cal{M}}_z)^2=-1$ and $({\mathcal{PT}})^2=-1$.
Nevertheless, one can use  $C_{2z}{\cal{T}}={\cal{M}}_z{\mathcal{PT}}$ instead of  ${\mathcal{PT}}$ to protect the  real topology  in 2D spinful systems, as  $C_{2z}{\cal{T}}$ not only keeps the 2D momentum  ${\bm k}=\{k_x,k_y\}$ invarient and always  satisfies $(C_{2z}{\cal{T}})^2=1$ regardless of the strength of the  SOC effect.

Similarly, the  definition of MRCI in spinful systems requires each mirror subsystem to  be invarient under $C_{2z}{\cal{T}}$.
However, this symmetry requirement is not satisfied generally.
For eigenstates $|\psi_{\pm}\rangle$ with ${\cal{M}}_z|\psi_{\pm}\rangle=\pm i|\psi_{\pm}\rangle$, one has
\begin{eqnarray}\label{eq:13}
{\cal{M}}_zC_{2z}{\cal{T}}|\psi_{\pm}\rangle &=& C_{2z}{\cal{T}}{\cal{M}}_z|\psi_{\pm}\rangle=C_{2z}{\cal{T}}(\pm i|\psi_{\pm}\rangle) \nonumber \\
&=&\mp i (C_{2z}{\cal{T}})|\psi_{\pm}\rangle),
\end{eqnarray}
indicating that under the action of $C_{2z}{\cal{T}}$, the eigenstates with ${\cal{M}}_z=\pm i$ are transformed into that with ${\cal{M}}_z=\mp i$.
Thus, it is the combination of these two mirror subsystems that makes  the entire system to have  $C_{2z}{\cal{T}}$, and each mirror subsystem breaks $C_{2z}{\cal{T}}$.

To address the above symmetry dilemma, we use projective symmetry operators  to alter  the   algebra of   $C_{2z}{\cal{T}}$ (${\mathcal{PT}}$)  and   ${\cal{M}}_z$.
For 2D spinful systems, the desired projective symmetry operators may be realized by choosing suitable  gauge field, which imposes important constraint on the phase of the hopping amplitudes \cite{shaoSpinlessMirrorChern2023}.
The $\mathbb{Z}_2$ gauge field introduces an additional degree of freedom: a sign of $+$ or $-$ to each  basis \cite{shaoSpinlessMirrorChern2023}.

We use a concrete lattice model to illustrate the influence  of the $\mathbb{Z}_2$ gauge field on the symmetry algebra.
Consider a 2D lattice model defined on  a square lattice, as shown in Fig. \ref{fig5}(a).
Here each unit cell contains two active site located at $\{(0,0,\frac{1}{4}),(0,0,-\frac{1}{4})\}$ position labeled as $A$ and $B$ respectively. At each site, we put four basis orbitals  $\Psi=\{|d_{z^{2}}\uparrow \rangle, |s\uparrow \rangle,|s\downarrow \rangle, |d_{z^{2}}\downarrow \rangle\}$.
Besides, the sites $A$ and $B$ are respectively endowed with a sign of $+$ or $-$ under a gauge transformation operator $\mathsf{G}$. After we arranged the following orbital sequence: $\{A|d_{z^{2}}\uparrow \rangle, A|s\uparrow \rangle,A|d_{z^2}\downarrow \rangle, A|s\downarrow \rangle, B|s\uparrow \rangle, B|d_{z^{2}}\uparrow \rangle,B|s\downarrow \rangle, B|d_{z^{2}}\downarrow \rangle\ \}$, we can get
\begin{eqnarray}\label{eq:14}
\mathsf{G} & = & \sigma_{3}\Gamma_{0,0}.
\end{eqnarray}
Here,  $\Gamma_{0,0}$ acts on the basis orbitals $\Phi$ and $\sigma_{3}$ acts on the site space.
Under the basis, the matrix representations of  ${\mathcal{M}}_z$ and ${\mathcal{PT}}$ are expressed as
\begin{eqnarray}\label{eq:15}
{\mathcal{M}}_z=i \sigma_{1}\Gamma_{3,1}, \ &  & {\mathcal{PT}}=-i\sigma_{1}\Gamma_{2,1}{\cal{K}} .
\end{eqnarray}
And further, the system breaks ${\cal{M}}_z$ and ${\mathcal{PT}}$ after adding the gauge, but exhibits the  combination operators $\mathsf{G}{\cal{M}}_z$ and $\mathsf{G}{\mathcal{PT}}$.
Since the gauge transformation operator $\mathsf{G}$ does not changes either site position in real space or ${\bm k}$ in momentum space \cite{shaoSpinlessMirrorChern2023}, ${\cal{M}}_z^{\prime}=\mathsf{G}{\cal{M}}_z$ and ${\cal{P}}^{\prime}\mathcal{T}=\mathsf{G}{\mathcal{PT}}$ can be considered as effective mirror and space-time inversion symmetries. Now we have
\begin{eqnarray}\label{eq:16}
{\cal{M}}_z^{\prime}=- \sigma_{2}\Gamma_{3,1}, \ &  & {{\cal{P}}^{\prime}\mathcal{T}}=\sigma_{2}\Gamma_{2,1} {\cal{K}}.
\end{eqnarray}
This means that one still can divide the system into two mirror subsystems based on the eigenvalue of  ${\cal{M}}_z^{\prime}$.
Particularly, due to the gauge field, one has $({\cal{M}}_z^{\prime})^2=1$, $({{\cal{P}}^{\prime}\mathcal{T}})^2=1$, so we can still use ${{\cal{P}}^{\prime}\mathcal{T}}$ to protect the real topology. Besides,
\begin{eqnarray}\label{eq:17}
{\cal{M}}_z^{\prime}{{\cal{P}}^{\prime}\mathcal{T}}|\tilde{\psi}_{\pm}\rangle &=& {{\cal{P}}^{\prime}\mathcal{T}}{\cal{M}}_z^{\prime}|\tilde{\psi}_{\pm}\rangle = {{\cal{P}}^{\prime}\mathcal{T}}(\pm 1|\tilde{\psi}_{\pm}\rangle) \nonumber \\
&=&\pm 1 ({{\cal{P}}^{\prime}\mathcal{T}}|\tilde{\psi}_{\pm}\rangle),
\end{eqnarray}
for the  ${\cal{M}}_z^{\prime}$'s eigenstates  $|\tilde{\psi}_{\pm}\rangle$ with ${\cal{M}}_z^{\prime}|\tilde{\psi}_{\pm}\rangle=\pm 1|\tilde{\psi}_{\pm}\rangle$.
Therefore, each mirror subsystem can be characterized by a real  Chern number and the total system features a  $\mathbb{Z}_2\oplus\mathbb{Z}_2$  topological index $\nu_M=(\nu_R^{+},\nu_R^{-})$.

The  lattice model that satisfies the  symmetries of ${\cal{M}}_z^{\prime} $ and $ {{\cal{P}}^{\prime}\mathcal{T}} $ may be written as
\begin{eqnarray}\label{eq:18}
H_{sf}^{2D}(\textbf{k}) & = &
\chi_{c}^{+}(\textbf{k})\sigma_{0}\Gamma_{0,0}+\chi_{s}^{-}(\textbf{k})\sigma_{0}\Gamma_{0,1} \nonumber \\
& & +\chi_{c}^{-}(\textbf{k})\sigma_{3}\Gamma_{0,3}-\chi_{s}^{+}(\textbf{k})\sigma_{1}\Gamma_{0,3}\nonumber \\
& & +(-m_{1}\sigma_{3}\Gamma_{0,0}+m_{2}\sigma_{0}\Gamma_{1,1}).
\end{eqnarray}
Here, $\chi_{c}^{\pm}(\textbf{k})=a_{1}\pm a_{2}+(t_{2}\pm t_{3})(\cos\,k_{x}+\cos\,k_{y}) $ and $\chi_{s}^{\pm}(\textbf{k})=t_{1}(\pm \sin\,k_{x}+\sin\,k_{y}) $. In this model, we only consider hoppings between the same spin. The energy band of this model (\ref{eq:18}) with $a_{1}=0.55,\,a_{2}=-0.45,\,t_{1}=1.2,\,t_{2}=0.4,\,t_{3}=-0.45 $  is plotted in Fig.  \ref{fig5}(c).

We then use the Wilson loop to calculate the real topology of the ${\cal{M}}_z^{\prime}=\pm1$ subsystem, and the obtained result for  ${\cal{M}}_z^{\prime}=-1$ subsystem is  shown  in Fig. \ref{fig5}(d), which  presents a single crossing at the lines of both $\theta=0$ and $\theta=\pi$. It means that $\nu_R^{-}=1$.
The result for the  ${\cal{M}}_z^{\prime}=1$ subsystem is the same.
This  clearly  demonstrates the existence  of a MRCI with $(\nu_R^{+},\nu_R^{-})=(1,1)$ in spinful systems.

Due to the rapid development in artificial periodic systems, the $\mathbb{Z}_2$ gauge field  is readily realized in acoustic crystals \cite{xue_observation_2020,qi_acoustic_2020,ni_demonstration_2020,xue_projectively_2022,li_acoustic_2022}.
However,  its realization  in  spinful systems is  still challenging.

\begin{figure}[t]
	\includegraphics[width=1\columnwidth]{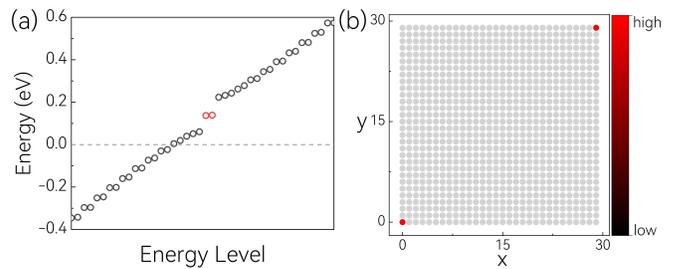}
	   \caption{(a) Energy spectrum of a $30\times30$ supercell for the ${\mathcal{M}}_{z}^{\prime}=-1$ subsystem, corner states are depicted by red circles. (b) The real-space distribution of the corner states marked by red circles in (a). }
	\label{fig6}
\end{figure}

\subsection{3D MRCIs}

It is known that the nonsymmorphic operators  can lead to completely different symmetry algebras inside the BZ and on the BZ boundary.
Then, one may think that with  certain nonsymmorphic operators, the mirror-invariant plane can be a MRCI even in the spinful systems without introducing the  gauge field discussed in Sec. \ref{2Dspinful}.
Unfortunately, we find that this is impossible.
For the space-time reversal symmetry $\widetilde{{\cal{PT}}}=\{{\cal{PT}}|\vec{\tau}\}$ with $\vec{\tau}$ a general fractional translation, one always has
\begin{eqnarray}\label{pt}
\widetilde{{\cal{PT}}}^2 &=& {(\cal{PT})}^2=-1.
\end{eqnarray}

Similarly, for a mirror symmetry $\widetilde{{\cal{M}}}_z=\{{\cal{M}}_z|\vec{\tau}\}$,
it also has $\widetilde{{\cal{M}}}_z^2=-1$, which indicates that
$\widetilde{{\cal{M}}}_z=\pm i$ for a generic point on both $k_{z}=0$ and $k_{z}=\pi$ planes.
Hence, to realize a MRCI on the mirror-invariant plane, two conditions must be satisfied:
\begin{eqnarray}\label{eq:21}
\widetilde{{C_{2z}\cal{T}}}^2=1,\  \{\widetilde{{\cal{M}}}_z,\widetilde{{C_{2z}\cal{T}}}\}=0,
\end{eqnarray}
if the Eq. (\ref{eq:21}) is satisfied, we have the follow equation:
 \begin{eqnarray}
\widetilde{{\cal{PT}}}^2 &=& (-\widetilde{{\cal{M}}}_z\widetilde{{C_{2z}\cal{T}}})(-\widetilde{{\cal{M}}}_z\widetilde{{C_{2z}\cal{T}}})
\nonumber \\
&=&-\widetilde{{\cal{M}}}_z^2(\widetilde{{C_{2z}\cal{T}}})^2=1,
\end{eqnarray}
where $\widetilde{{\cal{PT}}}=-\widetilde{{\cal{M}}}_z\widetilde{{C_{2z}\cal{T}}}$
, which is satisfied when considering SOC,  is used.
However, this above result  contradicts Eq. (\ref{pt}).
Thus, it is impossible  to achieve a MRCI in 3D real materials with strong SOC effect by the way of introducing the nonsymmorphic operators.

\subsection{Second-order boundary modes}

The second-order boundary modes in spinful systems is similar to that in spinless systems.
As mentioned, the lattice model (\ref{eq:18}) have $\nu_{M}=(\nu_R^{+},\nu_R^{-})=(1,1)$. Thus, it should have two sets of  corner states.
Based on the lattice model (\ref{eq:18}), we calculate  the energy spectrum of a $30\times30$ supercell for the ${\mathcal{M}}_{z}^{\prime}=-1$ subsystem, and the results are presented in Fig. \ref{fig6}(a).
One observes a pair of isolated states appearing in the bulk band gap, which are connected by ${{\mathcal{P}}^{\prime}\mathcal{T}}$. The real-space distribution of these states is depicted in Fig. \ref{fig6}(b), confirming that they are
the corner states.

\section{CONCLUSION AND DISCUSSION} \label{discussion}
In this paper, we theoretically establish the existence of symmetry-protected MRCI. For both 2D and 3D systems, with or without spin-orbit coupling, we offer theoretical analysis and detailed model construction. However, discovering materials with these topological states remains challenging, and our analysis offers guidance to this search.


In conclusion, we propose a new topological state,  named MRCI, protected by $\mathcal{PT}$ and ${\mathcal{M}}_{z}$ symmetries. We provide sufficient symmetry analysis and construct specific models for 2D and 3D systems, both with and without SOC, demonstrating their properties. Our findings show that a system with ${\mathcal{M}}_{z}$ and $\mathcal{PT}$ symmetries whose real Chern number is trivial may still have non-trivial topological properties, i.e. the real Chern number of its mirror subsystem may be nontrivial. In 2D systems with SOC, we address the problem of the absence of $\mathcal{PT}$ in subsystems using a $\mathbb{Z}_{2}$ gauge field. For 3D spinless systems, nonsymmorphic operations link two mirror subsystems, changing the classification of MRCIs from $\mathbb{Z}_2\oplus\mathbb{Z}_2$  classification to $\mathbb{Z}_2$ classification. For 3D spinful systems, we prove that it is not possible to realize MRCIs by introducing non-symmorphic operations. Across all scenarios, we identify second-order topological states associated with non-trivial MRCIs, such as corner states and hinge states. Our work enhances the understanding of real topologies and provides directions for realizing MRCIs in spinful systems and discovering new topological materials.

\begin{acknowledgments}
This work was supported by the NSF of China (Grants No. 12274028, No. 52161135108, No.12004035, and No. 12234003) and the National Key R\&D Program of China with Grant No. 2022YFA1402603.
\end{acknowledgments}


\bibliography{ref}

\end{document}